\documentstyle[amsmath,amssymb,graphicx]{article}

\def\be{\begin{eqnarray}}
\def\ee{\end{eqnarray}}
\def\nn{\nonumber}

\def\l[{\phantom.[}

\hoffset= - 1.1in         

\begin{document}

\hfill ITEP/TH-9/15

\hfill IITP/TH-3/15

\bigskip

\centerline{\Large{
On the {\it defect} and stability of differential expansion
}}

\bigskip

\centerline{{\bf Ya.Kononov$^d$, A.Morozov$^{a,b,c}$}}

\bigskip

{\footnotesize
\centerline{{\small
$^a$ ITEP, Moscow 117218, Russia}}

\centerline{{\small
$^b$ National Research Nuclear University MEPhI, Moscow 115409, Russia
}}

\centerline{{\small
$^c$ Institute for Information Transmission Problems, Moscow 127994, Russia
}}

\centerline{{\small
$^d$ Higher School of Economics, Math Department, Moscow, 117312, Russia
}}
}

\bigskip

\bigskip

\centerline{ABSTRACT}

\bigskip

{\footnotesize
Empirical analysis of many colored knot polynomials,
made possible by recent computational advances in Chern-Simons theory,
reveals their stability: for any given negative $N$ and any given knot the
set of coefficients of the polynomial in $r$-th symmetric representation
does not change with $r$, if it is large enough.
This fact reflects the non-trivial and previously unknown properties
of the differential expansion,
and it turns out that from this point of view there are universality classes
of knots, characterized by a single integer, which we call
{\it defect}, and which is in fact related to the power of Alexander polynomial.
}

\bigskip

\bigskip

HOMFLY polynomials are Wilson-loop averages in $3d$ Chern-Simons theory \cite{CS},
which in this simplest model depend only on the topology of the Wilson line (knot).
Therefore one can separate and study the group-theory properties
of observables -- and this is a non-trivial and very interesting problem,
for a brief summary of results see \cite{MMknots}.
From the quantum field theory perspective knot polynomials are direct generalization
of conformal blocks, and this relation \cite{Wit} provides one of the effective calculational
methods in knot theory.

Recent advances in \cite{gmmms,mmmrs}, based on the previous
considerations in \cite{Wit}-\cite{germs},
provided a way to systematically calculate
simplest colored HOMFLY polynomials \cite{knotpols} for a really wide variety of knots --
including, in particular, the entire Rolfsen table of \cite{katlas}.
This allows us to return to the study of "differential expansions" of
\cite{DGR}-\cite{Sle}, which was temporarily postponed
because of the insufficient "experimental" material.

In this note we describe empirically obtained properties of these expansions
for symmetric representations $[r]$ (where $r$ is the length of the single-line Young diagram).
It looks like there are different universality classes of such expansions,
characterized by a single integer, which we call "defect" $\delta^{\cal K}$.
Moreover, these newly observed properties allow to identify $2(\delta^{\cal K}+1)$ with
the power of Alexander polynomial and lead to a peculiar {\it stability} property of
symmetrically colored HOMFLY for large enough $r$:
what stabilizes is {\it not} the polynomial itself, but the set of its coefficients --
i.e. something like the "coordinates" $g_{r,j}$, introduced in \cite{arthdiff}.
Theoretical analysis of these  observations, proofs  and extension to non-(anti)symmetric
representations are beyond the scope of the present text.

\section{The notion of {\it defect}}

Differential expansion provides a  knot-dependent $q$-deformation (quantization) of the remarkable
factorization property \cite{LP,DMMMS},
\cite{IMMMfe}-\cite{Anton} of colored "special" polynomials at $q=1$,
\be
 \left.  H_R^{\cal K}(A)=\Big(H_{[1]}^{\cal K}(A)\Big)^{|R|}\right|_{q=1} \ \ \ \ \ \ \ \ \
 \forall \ \ \text{representation}\ R\ \text{and knot}\  {\cal K}
\label{spepoid}
\ee
which fully defines their dependence on representation (Young diagram) $R$.
Currently these expansions can be well
studied only for {\it symmetrically}-colored HOMFLY, and we focus on this case
in the present paper.

The story starts from the fact that

$\bullet$
$H_r = H_{[r]}$ always possesses differential expansion of the following form:
\be
H_r^{\cal K}(A,q^2) = 1 + \sum_{s=1}^r \frac{[r]!}{[s]![r-s]!}\cdot G_s^{\cal K}(A,q)
\cdot\{A/q\}\cdot \prod_{j=0}^{s-1} \{Aq^{r+j}\}
\label{diffe}
\ee
For generic knot $G_s^{\cal K}$ is a non-factorizable Laurent polynomial of $A$ and $q$,
but for some knots it can be further factorized.
In this formula we use the
notation $\{x\}=x-x^{-1}$ and quantum number
is defined as $[n] = \{q^n\}/\{q\}$.

$\bullet$
What is important, if $G_s^{\cal K}$ is divisible by some "differential" $\{Aq^k\}$,
the same is true for all other $G_{s'}^{\cal K}$ with $s'>s$.
This property allows one to introduce defect functions $\nu_s^{\cal K}$ and
$\mu_s^{\cal K}=s-1-\nu_s^{\cal K}$:
\be
G_s^{\cal K}(A,q) =
F_s^{\cal K}(A,q)\cdot \prod_{j=0}^{ \nu_s^{\cal K}-1}\{Aq^j\} =
F_s^{\cal K}(A,q)\cdot \prod_{j=0}^{s-2-\mu_s^{\cal K}}\{Aq^j\}
\ee
which are both(!) monotonically increasing function of $s$,
\be
\nu_s^{\cal K}\leq\nu_{s'}^{\cal K},\ \ \ \ \
\mu_s^{\cal K}\leq\mu_{s'}^{\cal K} \ \ \ \ \ {\rm for} \ \ \ \ \ s<s'
\ee
i.e. both grow -- but not faster than $s$.

$\bullet$
For $A=q^N$ with any fixed $N$, positive or negative,
\be
F_s(q^N,q)\sim \{q\}^{\mu_s^{\cal K}} \ \ \ \Longleftrightarrow \ \ \
G_s(q^N,q)\sim \{q\}^{s-1}
\label{degsmall}
\ee
i.e. at fixed $N$ the $s$-the term of differential expansion is actually
of the order $\{q\}^{2s}$.

$\bullet$
It turns out that  $\nu_s^{\cal K}$ as a function of $s$ has a very special shape,
fully  parameterized by a single integer
$\delta^{\cal K}\geq -1$, which we call the {\it defect} of differential expansion:

\be
\text{defect}\ \delta^{\cal K}=-1 \ \ \ \Longrightarrow \ \ \
\mu_s^{\cal K}=s-2, \ \ \ \ \nu_s^{\cal K}=1
\label{deltam1}
\ee

\be
\text{defect}\ \delta^{\cal K}=0 \ \ \ \Longrightarrow \ \ \ \mu_s^{\cal K}=0, \ \ \ \ \nu_s^{\cal K} = s-1
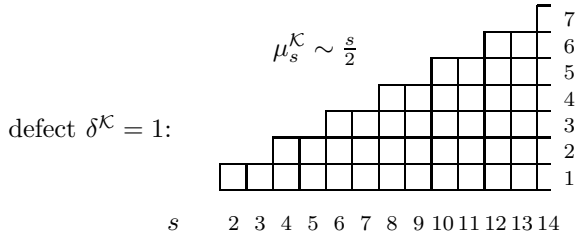
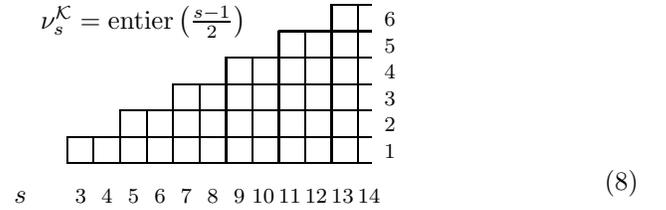
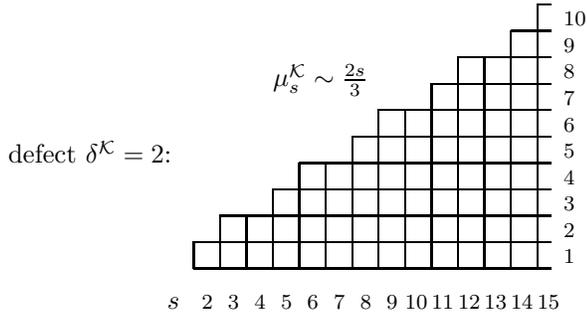
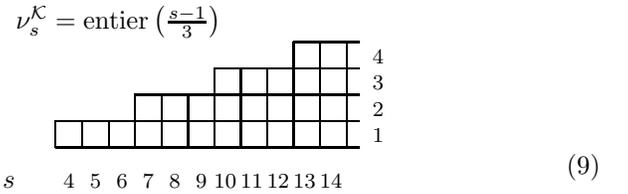
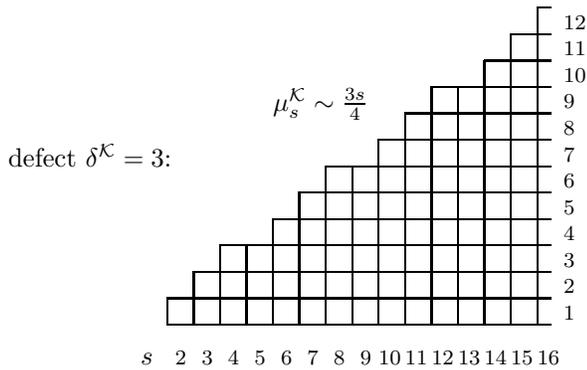
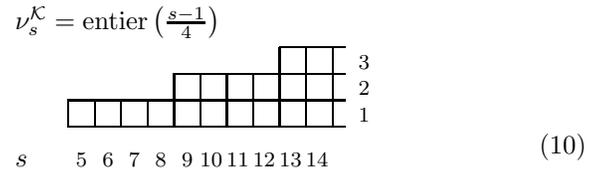
\ee

\be
\begin{picture}(300,70)(0,-30)
\put(-90,0){\mbox{defect $\delta^{\cal K}=1$:}}
\put(-10,-20){\line(1,0){125}}
\put(-10,-10){\line(1,0){125}}
\put(10,0){\line(1,0){105}}
\put(30,10){\line(1,0){85}}
\put(50,20){\line(1,0){65}}
\put(70,30){\line(1,0){45}}
\put(90,40){\line(1,0){25}}
\put(110,50){\line(1,0){5}}
\put(-10,-20){\line(0,1){10}}
\put(0,-20){\line(0,1){10}}
\put(10,-20){\line(0,1){20}}
\put(20,-20){\line(0,1){20}}
\put(30,-20){\line(0,1){30}}
\put(40,-20){\line(0,1){30}}
\put(50,-20){\line(0,1){40}}
\put(60,-20){\line(0,1){40}}
\put(70,-20){\line(0,1){50}}
\put(80,-20){\line(0,1){50}}
\put(90,-20){\line(0,1){60}}
\put(100,-20){\line(0,1){60}}
\put(110,-20){\line(0,1){70}}
\put(-30,-35){\mbox{$s$}}
\put(-7,-35){\mbox{{\footnotesize $2$}}}
\put(3,-35){\mbox{{\footnotesize $3$}}}
\put(13,-35){\mbox{{\footnotesize $4$}}}
\put(23,-35){\mbox{{\footnotesize $5$}}}
\put(33,-35){\mbox{{\footnotesize $6$}}}
\put(43,-35){\mbox{{\footnotesize $7$}}}
\put(53,-35){\mbox{{\footnotesize $8$}}}
\put(63,-35){\mbox{{\footnotesize $9$}}}
\put(70,-35){\mbox{{\footnotesize $10$}}}
\put(80,-35){\mbox{{\footnotesize $11$}}}
\put(90,-35){\mbox{{\footnotesize $12$}}}
\put(100,-35){\mbox{{\footnotesize $13$}}}
\put(110,-35){\mbox{{\footnotesize $14$}}}
\put(120,-18){\mbox{{\footnotesize $1$}}}
\put(120,-8){\mbox{{\footnotesize $2$}}}
\put(120,2){\mbox{{\footnotesize $3$}}}
\put(120,12){\mbox{{\footnotesize $4$}}}
\put(120,22){\mbox{{\footnotesize $5$}}}
\put(120,32){\mbox{{\footnotesize $6$}}}
\put(120,42){\mbox{{\footnotesize $7$}}}
\put(10,31){\mbox{$\mu_s^{\cal K}\sim \frac{s}{2}$}}
\put(200,-20){\line(1,0){115}}
\put(200,-10){\line(1,0){115}}
\put(220,0){\line(1,0){95}}
\put(240,10){\line(1,0){75}}
\put(260,20){\line(1,0){55}}
\put(280,30){\line(1,0){35}}
\put(300,40){\line(1,0){15}}
\put(200,-20){\line(0,1){10}}
\put(210,-20){\line(0,1){10}}
\put(220,-20){\line(0,1){20}}
\put(230,-20){\line(0,1){20}}
\put(240,-20){\line(0,1){30}}
\put(250,-20){\line(0,1){30}}
\put(260,-20){\line(0,1){40}}
\put(270,-20){\line(0,1){40}}
\put(280,-20){\line(0,1){50}}
\put(290,-20){\line(0,1){50}}
\put(300,-20){\line(0,1){60}}
\put(310,-20){\line(0,1){60}}
%
\put(180,-35){\mbox{$s$}}
\put(203,-35){\mbox{{\footnotesize $3$}}}
\put(213,-35){\mbox{{\footnotesize $4$}}}
\put(223,-35){\mbox{{\footnotesize $5$}}}
\put(233,-35){\mbox{{\footnotesize $6$}}}
\put(243,-35){\mbox{{\footnotesize $7$}}}
\put(253,-35){\mbox{{\footnotesize $8$}}}
\put(263,-35){\mbox{{\footnotesize $9$}}}
\put(270,-35){\mbox{{\footnotesize $10$}}}
\put(280,-35){\mbox{{\footnotesize $11$}}}
\put(290,-35){\mbox{{\footnotesize $12$}}}
\put(300,-35){\mbox{{\footnotesize $13$}}}
\put(310,-35){\mbox{{\footnotesize $14$}}}
\put(320,-18){\mbox{{\footnotesize $1$}}}
\put(320,-8){\mbox{{\footnotesize $2$}}}
\put(320,2){\mbox{{\footnotesize $3$}}}
\put(320,12){\mbox{{\footnotesize $4$}}}
\put(320,22){\mbox{{\footnotesize $5$}}}
\put(320,32){\mbox{{\footnotesize $6$}}}
%
\put(190,31){\mbox{$\nu_s^{\cal K} = \text{entier}\left( \frac{s-1}{2}\right)$}}
\end{picture}
\ee

\be
\begin{picture}(300,100)(0,-30)
\put(-90,20){\mbox{defect $\delta^{\cal K}=2$:}}
\put(-20,-20){\line(1,0){135}}
\put(-20,-10){\line(1,0){135}}
\put(-10,0){\line(1,0){125}}
\put(10,10){\line(1,0){105}}
\put(20,20){\line(1,0){95}}
\put(40,30){\line(1,0){75}}
\put(50,40){\line(1,0){65}}
\put(70,50){\line(1,0){45}}
\put(80,60){\line(1,0){35}}
\put(100,70){\line(1,0){15}}
\put(110,80){\line(1,0){5}}
%
\put(-20,-20){\line(0,1){10}}
\put(-10,-20){\line(0,1){20}}
\put(0,-20){\line(0,1){20}}
\put(10,-20){\line(0,1){30}}
\put(20,-20){\line(0,1){40}}
\put(30,-20){\line(0,1){40}}
\put(40,-20){\line(0,1){50}}
\put(50,-20){\line(0,1){60}}
\put(60,-20){\line(0,1){60}}
\put(70,-20){\line(0,1){70}}
\put(80,-20){\line(0,1){80}}
\put(90,-20){\line(0,1){80}}
\put(100,-20){\line(0,1){90}}
\put(110,-20){\line(0,1){100}}
\put(-30,-35){\mbox{$s$}}
\put(-17,-35){\mbox{{\footnotesize $2$}}}
\put(-7,-35){\mbox{{\footnotesize $3$}}}
\put(3,-35){\mbox{{\footnotesize $4$}}}
\put(13,-35){\mbox{{\footnotesize $5$}}}
\put(23,-35){\mbox{{\footnotesize $6$}}}
\put(33,-35){\mbox{{\footnotesize $7$}}}
\put(43,-35){\mbox{{\footnotesize $8$}}}
\put(53,-35){\mbox{{\footnotesize $9$}}}
\put(60,-35){\mbox{{\footnotesize $10$}}}
\put(70,-35){\mbox{{\footnotesize $11$}}}
\put(80,-35){\mbox{{\footnotesize $12$}}}
\put(90,-35){\mbox{{\footnotesize $13$}}}
\put(100,-35){\mbox{{\footnotesize $14$}}}
\put(110,-35){\mbox{{\footnotesize $15$}}}
\put(120,-18){\mbox{{\footnotesize $1$}}}
\put(120,-8){\mbox{{\footnotesize $2$}}}
\put(120,2){\mbox{{\footnotesize $3$}}}
\put(120,12){\mbox{{\footnotesize $4$}}}
\put(120,22){\mbox{{\footnotesize $5$}}}
\put(120,32){\mbox{{\footnotesize $6$}}}
\put(120,42){\mbox{{\footnotesize $7$}}}
\put(120,52){\mbox{{\footnotesize $8$}}}
\put(120,62){\mbox{{\footnotesize $9$}}}
\put(120,72){\mbox{{\footnotesize $10$}}}
\put(10,49){\mbox{$\mu_s^{\cal K}\sim \frac{2s}{3}$}}
\put(210,-20){\line(1,0){115}}
\put(210,-10){\line(1,0){115}}
\put(240,0){\line(1,0){85}}
\put(270,10){\line(1,0){55}}
\put(300,20){\line(1,0){25}}
%
\put(210,-20){\line(0,1){10}}
\put(220,-20){\line(0,1){10}}
\put(230,-20){\line(0,1){10}}
\put(240,-20){\line(0,1){20}}
\put(250,-20){\line(0,1){20}}
\put(260,-20){\line(0,1){20}}
\put(270,-20){\line(0,1){30}}
\put(280,-20){\line(0,1){30}}
\put(290,-20){\line(0,1){30}}
\put(300,-20){\line(0,1){40}}
\put(310,-20){\line(0,1){40}}
\put(320,-20){\line(0,1){40}}
\put(190,-35){\mbox{$s$}}
\put(213,-35){\mbox{{\footnotesize $4$}}}
\put(223,-35){\mbox{{\footnotesize $5$}}}
\put(233,-35){\mbox{{\footnotesize $6$}}}
\put(243,-35){\mbox{{\footnotesize $7$}}}
\put(253,-35){\mbox{{\footnotesize $8$}}}
\put(263,-35){\mbox{{\footnotesize $9$}}}
\put(270,-35){\mbox{{\footnotesize $10$}}}
\put(280,-35){\mbox{{\footnotesize $11$}}}
\put(290,-35){\mbox{{\footnotesize $12$}}}
\put(300,-35){\mbox{{\footnotesize $13$}}}
\put(310,-35){\mbox{{\footnotesize $14$}}}
\put(330,-18){\mbox{{\footnotesize $1$}}}
\put(330,-8){\mbox{{\footnotesize $2$}}}
\put(330,2){\mbox{{\footnotesize $3$}}}
\put(330,12){\mbox{{\footnotesize $4$}}}
%
\put(195,25){\mbox{$\nu_s^{\cal K}=\text{entier}\left( \frac{s-1}{3}\right)$}}
\end{picture}
\ee

\be
\begin{picture}(300,120)(0,-30)
\put(-90,40){\mbox{defect $\delta^{\cal K}=3$:}}
\put(-30,-20){\line(1,0){145}}
\put(-30,-10){\line(1,0){145}}
\put(-20,0){\line(1,0){135}}
\put(-10,10){\line(1,0){125}}
\put(10,20){\line(1,0){105}}
\put(20,30){\line(1,0){95}}
\put(30,40){\line(1,0){85}}
\put(50,50){\line(1,0){65}}
\put(60,60){\line(1,0){55}}
\put(70,70){\line(1,0){45}}
\put(90,80){\line(1,0){25}}
\put(100,90){\line(1,0){15}}
\put(110,100){\line(1,0){5}}
%
\put(-30,-20){\line(0,1){10}}
\put(-20,-20){\line(0,1){20}}
\put(-10,-20){\line(0,1){30}}
\put(0,-20){\line(0,1){30}}
\put(10,-20){\line(0,1){40}}
\put(20,-20){\line(0,1){50}}
\put(30,-20){\line(0,1){60}}
\put(40,-20){\line(0,1){60}}
\put(50,-20){\line(0,1){70}}
\put(60,-20){\line(0,1){80}}
\put(70,-20){\line(0,1){90}}
\put(80,-20){\line(0,1){90}}
\put(90,-20){\line(0,1){100}}
\put(100,-20){\line(0,1){110}}
\put(110,-20){\line(0,1){120}}
\put(-40,-35){\mbox{$s$}}
\put(-27,-35){\mbox{{\footnotesize $2$}}}
\put(-17,-35){\mbox{{\footnotesize $3$}}}
\put(-7,-35){\mbox{{\footnotesize $4$}}}
\put(3,-35){\mbox{{\footnotesize $5$}}}
\put(13,-35){\mbox{{\footnotesize $6$}}}
\put(23,-35){\mbox{{\footnotesize $7$}}}
\put(33,-35){\mbox{{\footnotesize $8$}}}
\put(43,-35){\mbox{{\footnotesize $9$}}}
\put(50,-35){\mbox{{\footnotesize $10$}}}
\put(60,-35){\mbox{{\footnotesize $11$}}}
\put(70,-35){\mbox{{\footnotesize $12$}}}
\put(80,-35){\mbox{{\footnotesize $13$}}}
\put(90,-35){\mbox{{\footnotesize $14$}}}
\put(100,-35){\mbox{{\footnotesize $15$}}}
\put(110,-35){\mbox{{\footnotesize $16$}}}
\put(120,-18){\mbox{{\footnotesize $1$}}}
\put(120,-8){\mbox{{\footnotesize $2$}}}
\put(120,2){\mbox{{\footnotesize $3$}}}
\put(120,12){\mbox{{\footnotesize $4$}}}
\put(120,22){\mbox{{\footnotesize $5$}}}
\put(120,32){\mbox{{\footnotesize $6$}}}
\put(120,42){\mbox{{\footnotesize $7$}}}
\put(120,52){\mbox{{\footnotesize $8$}}}
\put(120,62){\mbox{{\footnotesize $9$}}}
\put(120,72){\mbox{{\footnotesize $10$}}}
\put(120,82){\mbox{{\footnotesize $11$}}}
\put(120,92){\mbox{{\footnotesize $12$}}}
\put(10,61){\mbox{$\mu_s^{\cal K}\sim \frac{3s}{4}$}}
\put(220,-20){\line(1,0){105}}
\put(220,-10){\line(1,0){105}}
\put(260,0){\line(1,0){65}}
\put(300,10){\line(1,0){25}}
%
\put(220,-20){\line(0,1){10}}
\put(230,-20){\line(0,1){10}}
\put(240,-20){\line(0,1){10}}
\put(250,-20){\line(0,1){10}}
\put(260,-20){\line(0,1){20}}
\put(270,-20){\line(0,1){20}}
\put(280,-20){\line(0,1){20}}
\put(290,-20){\line(0,1){20}}
\put(300,-20){\line(0,1){30}}
\put(310,-20){\line(0,1){30}}
\put(320,-20){\line(0,1){30}}
\put(200,-35){\mbox{$s$}}
\put(223,-35){\mbox{{\footnotesize $5$}}}
\put(233,-35){\mbox{{\footnotesize $6$}}}
\put(243,-35){\mbox{{\footnotesize $7$}}}
\put(253,-35){\mbox{{\footnotesize $8$}}}
\put(263,-35){\mbox{{\footnotesize $9$}}}
\put(270,-35){\mbox{{\footnotesize $10$}}}
\put(280,-35){\mbox{{\footnotesize $11$}}}
\put(290,-35){\mbox{{\footnotesize $12$}}}
\put(300,-35){\mbox{{\footnotesize $13$}}}
\put(310,-35){\mbox{{\footnotesize $14$}}}
\put(330,-18){\mbox{{\footnotesize $1$}}}
\put(330,-8){\mbox{{\footnotesize $2$}}}
\put(330,2){\mbox{{\footnotesize $3$}}}
%
\put(200,17){\mbox{$\nu_s^{\cal K}=\text{entier}\left( \frac{s-1}{4}\right)$}}
\end{picture}
\ee

\be
\ldots
\nn
\ee

\noindent
In general
\be  \nu_s^{\cal K}=\text{entier}\left( \frac{s-1}{\delta^{\cal K}+1}\right)
\ \sim \  \frac{s}{\delta^{\cal K}+1},
\ \ \ \ \ \ \   \ \ \ \ \ \ \
\mu_s^{\cal K}= s-1-\nu_s^{\cal K} \ \sim \ \frac{\delta^{\cal K}}{\delta^{\cal K}+1}\cdot s
\ee

\section{Relation to Alexander polynomial}

It is an interesting question, if
the value of $\delta^{\cal K}$ can also restrict the coefficient functions $F_s^{\cal K}(A,q)$.

Immediately observable are two remarkable properties of this kind.

$\bullet$
$G_1^{\cal K}(A,q)$ has power $2\delta^{\cal K}$ in $q^2$, i.e. $G_1^{\cal K}(A,q)
= \sum\limits_{j=-\delta^{\cal K}}^{\delta^{\cal K}} c_jq^{2j}$.

\noindent
For example, $\delta^{\cal K}=0$ whenever $G_1^{\cal K}$ is independent of $q$.

$\bullet$ Alexander polynomial has power $2(\delta^{\cal K}+1)$  in $q^2$,
i.e. $Al^{\cal K}(q) = H_1^{\cal K}(A=1,q)
=\sum\limits_{j=-\delta^{\cal K}-1}^{\delta^{\cal K}+1} a_jq^{2j}$

\be
\delta^{\cal K} = \frac{1}{2}\text{Power}_{q^2}(Al^{\cal K}) - 1
\label{degAl}
\ee

For $\delta^{\cal K}\neq 0$ these facts are not immediately related:
contributing to Alexander polynomials are all $G_s^{\cal K}$ with $s\leq \delta^{\cal K}+1$
and they can and do contain much higher powers in $q$.
Moreover, even the product of differentials in the $s$-term has power in $q$, which grows
quadratically with $s$ -- and thus with $\delta^{\cal K}$.
This means that there are serious cancelations behind the linear law (\ref{degAl}).

Since Alexander polynomials are easily available already from \cite{katlas},
we can now list the defects of all the knots from the Rolfsen table
(up to 10 intersections):
\be
\!\!\!\!\!\!\!\!\!\!\!\!\!\!\!\!\!\!\!\!\!\!\!\!\!\!\!\!\!\!\!\!\!\!\!\!\!\!\!\!\!\!\!\!\!\!\!\!\!\!\!\!\!\!\!\!\!\!
\begin{array}{c||c|c|cc|ccc|ccccccc|ccccccccccccccccccccc}
{\cal K} &3_1 & 4_1 & 5_1 & 5_2 & 6_1& 6_2&6_3   &7_1&7_2&7_3&7_4&7_5&7_6&7_7
\\   \hline
\delta^{\cal K}&  0&0&1&0&0&1&1  &2&0&1&0&1&1&1
\end{array}
\nn \\
\nn \\
\!\!\!\!\!\!\!\!\!\!\!\!\!\!\!\!\!\!\!\!\!\!\!\!\!\!\!\!\!\!\!\!\!\!\!\!\!\!\!\!\!\!\!\!\!\!\!\!\!\!\!\!\!\!\!\!\!\!
\begin{array}{c|| cccccccccc|cccccccccc|c}
{\cal K}  &8_1&8_2&8_3&8_4&8_5&8_6&8_7&8_8&8_9&8_{10}&8_{11}&8_{12}&8_{13}&8_{14}&
8_{15}&8_{16}&8_{17}&8_{18}&8_{19}&8_{20}&8_{21}
\\   \hline
\delta^{\cal K}&0 & 2 &0&1&2&1&2&1&2&2&1&1&1&1&1&2&2&2&2&1&1
\end{array}
\nn \\
\nn \\
\!\!\!\!\!\!\!\!\!\!\!\!\!\!\!\!\!\!\!\!\!\!\!\!\!\!\!\!\!\!\!\!\!\!\!\!\!\!
\begin{array}{c|| cccccccccc|cccccccccc|ccccc}
{\cal K}  &9_1&9_2&9_3&9_4&9_5&9_6&9_7&9_8&9_9&9_{10}&9_{11}&9_{12}&9_{13}&9_{14}&
9_{15}&9_{16}&9_{17}&9_{18}&9_{19}&9_{20} &9_{21}&9_{22}& 9_{23}&9_{24} &9_{25}
\\   \hline
\delta^{\cal K}& 3&0&2&1&0&2&1&1&2&1&2&1&1&1&1&2&2&1&1&2&1&2&1&2&1
\end{array}
\nn \\
\nn \\
\!\!\!\!\!\!\!\!\!\!\!\!\!\!\!\!\!\!\!\!\!\!\!\!\!\!\!\!\!\!
\begin{array}{c|| ccccc|cccccccccc|ccccccccc}
{\cal K}  &9_{26}&9_{27}&9_{28}&9_{29}&9_{30}&9_{31}&9_{32}&9_{33}&9_{34}&9_{35}&
9_{36}&9_{37}&9_{38}&9_{39}&
9_{40}&9_{41}&9_{42}&9_{43}&9_{44}&9_{45} &9_{46}&9_{47}& 9_{48}&9_{49}
\\   \hline
\delta^{\cal K}& 2&2&2&2&2&2&2&2&2&0&2&1&1&1&2&
1&1&2&1&1&0&2&1&1
\end{array} \nn
\ee

\bigskip

\bigskip

\be
\!\!\!\!\!\!\!\!\!\!\!\!\!\!\!\!\!\!\!\!\!\!\!\!\!\!\!
\begin{array}{c|| cccccccccc|cccccccccc| }
{\cal K}  &10_1&10_2&10_3&10_4&10_5&10_6&10_7&10_8&10_9&10_{10}&10_{11}&10_{12}&10_{13}&10_{14}&
10_{15}&10_{16}&10_{17}&10_{18}&10_{19}&10_{20}
\\   \hline
\delta^{\cal K}&  0&3&0&1&3&2&1&2&3&1&1&2&1&2&2&1&3&1&2&1
\end{array}
\nn \\
\nn \\
\!\!\!\!\!\!\!\!\!\!\!\!\!\!\!\!\!\!\!\!\!\!\!\!\!\!\!\!\!\!
\begin{array}{c|| cccccccccc|cccccccccc| }
{\cal K} &10_{21}&10_{22}& 10_{23}&10_{24} &10_{25} &10_{26}&10_{27}&10_{28}&10_{29}&10_{30}&
10_{31}&10_{32}&10_{33}&10_{34}&10_{35}&10_{36}&10_{37}&10_{38}&10_{39}&10_{40}
\\   \hline
\delta^{\cal K}&2&2&2&1&2& 2&2&1&2&1&1&2&1&1&1&1&1&1&2&2
\end{array}
\nn \\
\nn \\
\!\!\!\!\!\!\!\!\!\!\!\!\!\!\!\!\!\!\!\!\!\!\!\!\!\!\!\!\!\!
\begin{array}{c|| cccccccccc|cccccccccc|}
{\cal K}  &10_{41}&10_{42}&10_{43}&10_{44}&10_{45} &10_{46}&10_{47}& 10_{48}&10_{49} &10_{50}
&10_{51}&10_{52}&10_{53}&10_{54}&10_{55} &10_{56}&10_{57}& 10_{58}&10_{59} &10_{60}
\\   \hline
\delta^{\cal K}   & 2&2&2&2&2&3&3&3&2&2 &3&3&2&3&2&3&3&2&3&3
\end{array}
\nn \\
\nn \\
\!\!\!\!\!\!\!\!\!\!\!\!\!\!\!\!\!\!\!\!\!\!\!\!\!\!\!\!\!\!
\begin{array}{c|| cccccccccc|cccccccccc| }
{\cal K}  &10_{61}&10_{62}&10_{63}&10_{64}&10_{65} &10_{66}&10_{67}& 10_{68}&10_{69} &10_{70}
&10_{71}&10_{72}&10_{73}&10_{74}&10_{75} &10_{76}&10_{77}& 10_{78}&10_{79} &10_{80}
\\   \hline
\delta^{\cal K}&2&3&1&3&2&2&1&1&2&2&2&2&2&1&2&2&2&2&3&2
\end{array}
\nn \\
\nn \\
\!\!\!\!\!\!\!\!\!\!\!\!\!\!\!\!\!\!\!\!\!\!\!\!\!\!\!\!\!\!
\begin{array}{c|| cccccccccc|cccccccccc| }
{\cal K}   &10_{81}&10_{82}&10_{83}&10_{84}&10_{85} &10_{86}&10_{87}& 10_{88}&10_{89} &10_{90}
&10_{91}&10_{92}&10_{93}&10_{94}&10_{95} &10_{96}&10_{97}& 10_{98}&10_{99} &10_{100}
\\   \hline
\delta^{\cal K}&2&3&2&2&3&2&2&2&2&2&3&2&2&3&2&2&1&2&3&3
\end{array}
\nn \\
\nn \\
\begin{array}{c|| cccccccccc|ccccc}
{\cal K}  &10_{101}&10_{102}&10_{103}&10_{104}&10_{105} &10_{106}&10_{107}& 10_{108}&10_{109} &10_{110}
&10_{111}&10_{112}&10_{113}&10_{114}&10_{115}
\\   \hline
\delta^{\cal K}& 1&2&2&3&2&3&2&2&3&2&2&3&2&2&2
\end{array}
\nn \\
\nn \\
\begin{array}{c|| ccccc|cccccccccc| }
{\cal K}  &10_{116}&10_{117}& 10_{118}&10_{119}
&10_{120}&10_{121}&10_{122}&10_{123}&10_{124}&10_{125} &10_{126}&10_{127}& 10_{128}&10_{129} &10_{130}
\\   \hline
\delta^{\cal K}&3&2&3&2&1& 2&2&3&3&2&2&2&2&1&1
\end{array}
\nn \\
\nn \\
\begin{array}{c|| cccccccccc|ccccc}
{\cal K}
&10_{131}&10_{132}&10_{133}&10_{134}&10_{135} &10_{136}&10_{137}& 10_{138}&10_{139} &10_{140}
&10_{141}&10_{142}&10_{143}&10_{144}&10_{145}
\\   \hline
\delta^{\cal K}&1&1&1&2&1&1&1&2&3&1& 2&2&2&1&1
\end{array}
\nn \\
\nn \\
\begin{array}{c|| ccccc|cccccccccc| }
{\cal K}   &10_{146}&10_{147}& 10_{148}&10_{149} &10_{150}
&10_{151}&10_{152}&10_{153}&10_{154}&10_{155}&10_{156}&10_{157}&10_{158}&10_{159}&
10_{160}
\\   \hline
\delta^{\cal K}&1&1&2&2&2&2&3&2&2&2&2&2&2&2&2
\end{array}
\nn \\
\nn \\
\begin{array}{c|| ccccc}
 {\cal K}       &10_{161}&10_{162}&10_{163}&10_{164}&10_{165}
 \\   \hline
\delta^{\cal K}   &2&1&2&1&1
\end{array}
\nn
\ee

\section{Twist and torus knots}

For all twist knots the defect is vanishing
\be
\delta^{\text{twist}} = 0
\ee

\noindent
Instead for torus knots it is a kind of {\it maximal}:
\be
\text{for the 2-strand family} \ \  &\delta^{[2,n]} = \frac{n-3}{2}, \nn \\
\text{for the 3-strand family} \ \ (8_{19},\ 10_{124}, \ \ldots)
&\delta^{[3,n]} = n-2, \nn \\
\text{for the 4-strand family} \ \  & \delta^{[4,n]} = \frac{3n-5}{2} \nn \\
\ldots \nn \\
\text{in general}, & \delta^{[m,n]} = \frac{mn-m-n-1}{2},
\ee
since the power of Alexander polynomial $Al^{[m,n]}$ is $(m-1)(n-1)$.

\section{Negative defect: KTC mutants and their relatives}

Starting from $11$ intersections there are cases when Alexander is just unity,
i.e. the defect is negative, $\delta^{\cal K}=-1$.
According to our general rules this means that for such knots already $G_1^{\cal K}$ is
reducible: $G_1^{\cal K}\sim \{A\}$.
Of course, also all other $G_s^{\cal K}\sim \{A\}$, because all the terms
of the differential expansion are vanishing for $A=1$.

This is indeed true for the first example -- the celebrated
 Kinoshita-Terasaka and Conway (KTC) mutants
${\cal K}=11n42\ \&\ 11n34$,
reconsidered recently in \cite{mmmrs},--
and also for the next example, available from \cite{katlas}:
${\cal K}= 12n313\ \&\ 12n430$.
Moreover,
the combination of
\cite{gmmms} and \cite{mmmrs}  allows to calculate
HOMFLY for KTC mutants for any symmetric representation and validate (\ref{deltam1})
in this particular example.

\section{Summary: {\it stability} and other properties of differential expansion}

To explain what we mean by {\it stability}, it is simplest to look at an example of a randomly chosen knot
(say, ${\cal K}=6_2$):

{\tiny
\be
\begin{array}{rl}
\!\!\!\!\!\!\!\!\!\!\!\!\!\!\!\!\!\!\!\!\!\!\!\!\!\!
H_1^{6_2}\!\left(\!A=\frac{1}{q}\right) = &1
\nn \\
\!\!\!\!\!\!\!\!\!\!\!\!\!\!\!\!\!\!\!\!\!\!\!\!\!\!
q^{10}\cdot H_2^{6_2}\!\left(\!A=\frac{1}{q}\right) = &1  -  3 q^2  +q^4+5 q^6 - 8 q^8 + 3 q^{10} +10 q^{12}
- 10 q^{14} - 4 q^{16} + 9 q^{18} - q^{20} - 3 q^{22} +q^{24}
\nn \\
\!\!\!\!\!\!\!\!\!\!\!\!\!\!\!\!\!\!\!\!\!\!\!\!\!\!
q^{18}\cdot H_3^{6_2}\!\left(\!A=\frac{1}{q}\right) = &1  -  3 q^4  -  3 q^6   + 4 q^{8} + 9 q^{10} - 2 q^{12}
-12 q^{14} - 6 q^{16}
+ 11 q^{18}+14 q^{20} -2 q^{22} -12  q^{24}- 8 q^{26}+ 4 q^{28}+ 9 q^{30}+2 q^{32}-3 q^{34}
-3 q^{36}+q^{40}
\nn \\
\!\!\!\!\!\!\!\!\!\!\!\!\!\!\!\!\!\!\!\!\!\!\!\!\!\!
q^{26}\cdot H_4^{6_2}\!\left(\!A=\frac{1}{q}\right) = &1  -  3 q^6  -  3 q^8   + 4 q^{12} + 9 q^{14} + 2 q^{16}-4 q^{18} - 12 q^{20}
-8 q^{22}+2 q^{24}+11 q^{26}+14 q^{28}+2 q^{30}-4 q^{32}-12 q^{34}-8 q^{36}+4 q^{40}
+9 q^{42}+2 q^{44}-3 q^{48}-3 q^{50}+q^{56}
\nn\\
\!\!\!\!\!\!\!\!\!\!\!\!\!\!\!\!\!\!\!\!\!\!\!\!\!\!
q^{34}\cdot H_5^{6_2}\!\left(\!A=\frac{1}{q}\right) = &1  -  3 q^8  -  3 q^{10} + 4 q^{16} + 9 q^{18} + 2 q^{20}-4 q^{24} - 12 q^{26}
-8 q^{28}+2 q^{32}+11 q^{34}+14 q^{36}+2 q^{38}-4 q^{42}-12 q^{44}-8 q^{46}+4 q^{52}+9 q^{54}+2 q^{56}
-3 q^{62}-3 q^{64}+q^{72}
\nn \\
\!\!\!\!\!\!\!\!\!\!\!\!\!\!\!\!\!\!\!\!\!\!\!\!\!\!
q^{42}\cdot H_6^{6_2}\!\left(\!A=\frac{1}{q}\right) = &1  -  3 q^{10}-  3 q^{12} + 4 q^{20} + 9 q^{22} + 2 q^{24}-4 q^{30} - 12 q^{32}
-8 q^{34}+2 q^{40}+11 q^{42}+14 q^{44}+2 q^{46}-4 q^{52}-12 q^{54}-8 q^{56}+4 q^{64}+9 q^{66}+2 q^{68}
-3 q^{76}-3 q^{78}+q^{88}
\nn \\                  \nn \\
\ldots \ \ \ \ \  &
\nn
\end{array}
\ee
}

It is easy to observe that, starting from $H_4^{6_2}$,
the {\it sets of coefficients} are the same -- despite the polynomials are {\it different}.
At $A=q^{-2}$ the same happens, beginning from $H_8^{6_2}$.
Thus what stabilizes are not the polynomials themselves, but something else, more appropriately
associated with the knots.
In full accordance with the vision in \cite{arthdiff} this {\it something} are the {\it coefficient
functions} $G_s^{\cal K}$ of the differential expansion.

Due to their properties, which are revealed in the present paper,
contributing at $A=q^{-N}$ are just the first few terms of the
expansion (\ref{diffe}):
\be
H_r^{\cal K}\!\left(A=\frac{1}{q^{N}}\right) \ = \  1 \  -  \sum_{s=0}^{(N+1)(\delta^{\cal K}+1)}\
\frac{[N+1]\cdot G_s^{\cal K}\!\left(A=q^{-N}\right)}{\{q\}^{s-1}\cdot [s]!}\
\prod_{j=0}^{s-1}\{q^{r-N+j}\}\{q^{r-j}\}
\ee
where the last product is Laurent polynomial in $q^r$ and
due to (\ref{degsmall}) the ratio in front of it is an $r$-independent polynomial.
Thus what we get is just a sum of a few {\it polynomials}, multiplied by different powers of $q^r$.
They do not overlap at large enough $r$, and
this provides an $r$-independent set of the coefficients, as in the above example.

In fact, one could wish to interpret the remarkable identity \cite{IMMMfe,Zhu}
\be
H_r^{\cal K}(A=1,q) \ = \ H_1^{\cal K}(A=1,q^r)
\label{Alid}
\ee
for Alexander polynomials as a manifestation of the same phenomenon at $N=0$.
However this is {\it literally} so only for $\delta^{\cal K}=-1$ and $\delta^{\cal K}=0$.
Still (\ref{Alid}) is true not only for {\it all knots}, but actually for {\it all single-hook}
(and not just single-line) representations $R$.
For such representations (\ref{Alid}) is a kind of a dual to (\ref{spepoid}).

\section{Conclusion}

In this paper we studied the "quality" of the differential expansion (\ref{diffe}) for
symmetrically colored reduced HOMFLY polynomials --
the typical observables in the simplest possible Yang-Mills theory.
If only naive representation-theory properties are taken into account
from (\ref{spepoid}) to restriction $l\leq N$ on the number $l$ of lines
in the Young diagram for particular $SL(N)$,
this expansion has the form (\ref{diffe}) with irreducible polynomial
coefficient functions $G_s(A,q)$.
It is well known, however, that sometime $G_s$ are further factorized,
thus adding more restrictions/structures to the color-dependence
of physical observables.
Now, when methods were developed to study entire classes
of generic knots, we could attack this problem in a systematic way --
and demonstrate that $G_s$ are {\it always} factorizable for high
enough $s$.
The depth of factorization appeared to depend on a single
characteristic of the knot, which we originally called {\it defect} of
the expansion, and further demonstrated that it is linearly
related to the degree of Alexander polynomial, what makes it
very easy to find.

This factorization universality leads to remarkable kind of stabilization
of symmetrically colored HOMFLY -- ensuring that increasing $r$
beyond some knot-dependent boundary does not provide
new physical (topological) information.
This is what one naturally {\it expects}, and now we see how this actually {\it works}.

Highly desirable is extension of this new insight beyond pure symmetric
and antisymmetric representations, but this requires further
development of technical tools in conformal, quantum group and
${\cal R}$-matrix theories.

\section*{Acknowledgements}

Our work is partly supported by grants NSh-1500.2014.2, RFBR 13-02-00478
and by the joint grants 15-52-50041-YaF, 14-01-92691-Ind-a, 15-51-52031-NSC-a.
Acknowledged is also support by  Brazilian Ministry of Science, Technology and Innovation
through the National Counsel of Scientific and Technological Development and
by Laboratories of Algebraic Geometry and Mathematical Physics, HSE.

\end{document}